\documentclass[preprint,aps,prd,superscriptaddress,preprintnumbers,nofootinbib]{revtex4-1}
\usepackage{graphicx} % Required for inserting images
\usepackage{amsfonts}
\usepackage{amsmath}
\usepackage{amsthm}
\usepackage{amsmath}
\usepackage{booktabs}
\usepackage{geometry}
\usepackage{amssymb}
\usepackage{array}
\usepackage{dcolumn}
\usepackage{appendix}
\usepackage[svgnames]{xcolor}
\usepackage[hidelinks]{hyperref}
\hypersetup{
    colorlinks=true,
    linkcolor=NavyBlue,
    filecolor=NavyBlue,
    urlcolor=NavyBlue,
    citecolor=NavyBlue,
    }

\begin{document}

\title{Constraining the pseudo-Dirac nature of neutrinos using astrophysical neutrino flavor data}

\author{Chee Sheng Fong}
\email{sheng.fong@ufabc.edu.br}
\author{Yago Porto}
\email{yago.porto@ufabc.edu.br}
\affiliation{Centro de Ci\^{e}ncias Naturais e Humanas$,$ Universidade Federal do ABC$,$ 09.210-170$,$ Santo Andr\'{e}$,$ SP$,$ Brazil}

\begin{abstract}
The three Standard Model neutrinos can have Majorana mass or strictly Dirac mass, but both scenarios are practically indistinguishable in neutrino oscillation experiments.
If they are pseudo-Dirac, however, there will be new mass splittings among the pseudo-Dirac pairs, potentially leaving traces in neutrino oscillation phenomena. In this work, we use flavor ratios of astrophysical neutrinos to discriminate different possible mass spectra of pseudo-Dirac neutrinos. We show that it will be possible to impose robust bounds of order $\delta m^2_3 \lesssim 10^{-12}$ $\text{eV}^2$ on the new mass squared splitting involving the third pseudo-Dirac mass eigenstates (those with the least electron flavor composition) with the future experiment IceCube-Gen2. The derived sensitivity is robust because it only assumes an extragalactic origin for the astrophysical neutrinos and hierarchical pseudo-Dirac mass spectrum. In case the neutrino sources are known in the future, such bounds can potentially improve by up to five orders of magnitude, reaching $\delta m^2_3 \lesssim 10^{-17}$ $\text{eV}^2$.
\end{abstract}

\maketitle

\newpage

\section{Introduction}

Nonzero neutrino mass with at least one mass eigenstate with mass
of greater than 0.05 eV has been established experimentally through
neutrino oscillation phenomena \cite{deSalas:2020pgw,Esteban:2020cvm}. In the Standard Model (SM),
neutrinos, being electrically neutral, are the only fermions that
can acquire Majorana mass without breaking the electromagnetic gauge
symmetry. To respect the full SM gauge symmetry, Majorana neutrino
mass can only arise from new physics at some scale $\Lambda$ which
breaks the the total lepton number by two units. 
This leads to neutrinoless double beta decay, which, in the case of an inverted neutrino mass ordering scenario, should be observable in future experiments such as nEXO~\cite{nEXO:2021ujk}.
On the other hand,
if neutrinos were to have Dirac mass, new fermion degrees of freedom
(the right-handed neutrinos) have to be introduced. In this case,
since lepton number is conserved, neutrinoless double beta decay signature
will be absent. In neutrino oscillation experiments where lepton number
is conserved, one cannot distinguish between Majorana and Dirac neutrinos.
However, if the lepton number is slightly broken such that neutrinos
are pseudo-Dirac, there will be new small mass splitting among the
pseudo-Dirac pair.\footnote{The same scenario is referred to as quasi-Dirac in refs. \cite{Anamiati:2017rxw,Anamiati:2019maf,Fong:2020smz}
but we will opt for pseudo-Dirac, which is a more popular choice in the literature.} While the rate of neutrinoless double beta decay will be suppressed
due to small lepton number violation, interestingly, one can probe
this scenario in neutrino oscillation experiments as we will discuss next.

In pseudo-Dirac scenario~\cite{Kobayashi:2000md}, we can have up to three pairs of mass eigenstates
($i=1,2,3$) with squared masses
\begin{eqnarray}
\hat{m}_{j}^{2} & = & m_{j}^{2}-\frac{1}{2}\delta m_{j}^{2},\qquad\hat{m}_{j+3}^{2}=m_{j}^{2}+\frac{1}{2}\delta m_{j}^{2}, \label{eq:new_squared_mass_splitting}
\end{eqnarray}
where $m_j^2$ are the standard mass squared eigenvalues and $\delta m_{j}^{2}$ are the three new mass squared splittings
which have not been observed experimentally and are expected to be
smaller than the solar mass splitting. Due to the pseudo-Dirac structure,
the mixing between the SM and the right-handed neutrinos are close
to maximal. Under the assumption of maximal mixing, strong experimental
constraints using solar neutrino data have been derived: $\delta m_{1,2}^{2}\lesssim10^{-11}\,\textrm{eV}^{2}$,
while it is not sensitive to $\delta m_{3}^{2}$ due to small $\theta_{13}$
\cite{Cirelli:2004cz,deGouvea:2009fp,Anamiati:2017rxw,Ansarifard:2022kvy}. (See also \cite{Franklin:2023diy}.)
Using atmospheric, beam and reactor neutrino data which are sensitive to
atmospheric mass splitting and $\theta_{13}$, a much weaker constraint
is obtained $\delta m_{3}^{2}\lesssim10^{-5}\,\textrm{eV}^{2}$ \cite{Anamiati:2017rxw}.
(See also \cite{Cirelli:2004cz}.) Due to the small mass splitting,
astrophysical neutrinos coming from distance sources of Mpc to Gpc
allow to probe mass squared difference much smaller than $10^{-12}\,\textrm{eV}^{2}$
\cite{Crocker:1999yw,Crocker:2001zs,Cirelli:2004cz,Esmaili:2009fk,Esmaili:2012ac,Joshipura:2013yba,Shoemaker:2015qul,Brdar:2018tce,DeGouvea:2020ang,Martinez-Soler:2021unz,Rink:2022nvw,Carloni:2022cqz,Dixit:2024ldv}. 
While the pseudo-Diracness of neutrinos are interesting in itself,
ref.~\cite{Fong:2020smz} has showed that it could be linked to the cosmic baryon
asymmetry for pseudo-Dirac seesaw model where the new squared mass splitting
is intimately connected to the CP violation $\epsilon$ required for baryogenesis
through leptogenesis as follows\footnote{The pseudo-Dirac leptogenesis model proposed in ref.~\cite{Ahn:2016hhq} does not impose such a strong constraint, predicting only a lower bound on seesaw scale for a given pseudo-Dirac mass splitting.}
\begin{eqnarray}
\delta m_{i}^{2} & \gtrsim & 10^{-8}\,\textrm{eV}^{2}\left(\frac{m_{i}}{0.1\,\mathrm{eV}}\right)^{2}\left(\frac{\epsilon}{10^{-7}}\right).
\end{eqnarray}
Here $\epsilon\sim10^{-7}$ represents roughly the minimum value
required to explain the observed baryon asymmetry. Given current constraints, this can only be achieved by $\delta m_3^2$. This provides further motivation to explore the third pseudo-Dirac mass splitting within this range.

The advent of neutrino telescopes made it possible to study neutrino properties using high-energy (HE) astrophysical neutrinos \cite{Pakvasa:2004hu}. Neutrino telescopes employ naturally occurring targets with large volumes to detect tiny astrophysical neutrino fluxes amidst the overwhelming backgrounds of atmospheric neutrinos and muons \cite{Arguelles:2024xkx}. The IceCube Neutrino Observatory, for instance, has utilized a cubic kilometer of Antarctic ice since 2011 to observe neutrinos in the TeV–PeV range \cite{Ahlers:2018fkn}. Following the discovery of HE neutrinos in 2013 \cite{IceCube:2013cdw, IceCube:2013low}, further observations by IceCube \cite{IceCube:2014stg, IceCube:2015gsk, IceCube:2015qii, IceCube:2016umi, IceCube:2020fpi, IceCube:2020wum, IceCube:2021uhz}, ANTARES \cite{ANTARES:2017srd}, and Baikal-GVD \cite{Baikal-GVD:2022fis} initiated an ongoing phase of characterizing the diffuse neutrino flux and searching for candidate neutrino sources \cite{IceCube:2018cha, IceCube:2018dnn, Rodrigues:2020fbu, IceCube:2022der, ANTARES:2023lck}. However, despite the recent identification of the first neutrino sources in the northern hemisphere, the origins of most of the diffuse flux are still unknown \cite{Kurahashi:2022utm, Troitsky:2023nli}, and the absence of strong anisotropies in the flux indicates an extragalactic origin, with a minor galactic contribution at the level of $10\%$ \cite{Ahlers:2015moa, IceCube:2023ame, Bustamante:2023iyn}.

The extragalactic origin of the HE neutrinos suggests that these particles travel, from their production to detection, a minimum distance of $L \approx 15$ kpc, equivalent to the radius of the Milky Way Galaxy. This estimate is notably conservative, considering that the Andromeda Galaxy, the closest major galaxy, lies 600 kpc away, while the nearest Active Galactic Nucleus (NGC5128) is situated roughly 4 Mpc from the Milky Way. This estimate is also compatible with the redshift evolution of BL Lac objects \cite{Ajello:2013lka, Capanema:2020oet}. 
The important quantity which allows pseudo-Dirac neutrinos to be constrained is the pseudo-Dirac oscillating phase $\Phi_j \sim \delta m_j^2 L/E$. If $\Phi_j \ll 1 \implies \delta m_j^2 \ll E/L$, the pseudo-Dirac pair remains coherent throughout the propagation and the oscillation approaches the standard scenario and we lose the sensitivity to constrain the scenario. If $\Phi_j \gg 1$, we have the \emph{decoherence} among the pseudo-Dirac pair and since sterile states are not detected, the corresponding probability amplitude is suppressed by a factor of two.
%However, even with such a conservative choice of baseline, 
Even with a conservative choice of $L \sim 15$ kpc, under the assumption of hierarchical pseudo-Dirac mass splittings,
we anticipate that IceCube-Gen2 \cite{IceCube-Gen2:2020qha} should be able to impose the constraint $\delta m_3^2 \lesssim 10^{-12}$ eV$^2$ due to its improved flavor ratio measurements~\cite{IceCube-Gen2:2023rds}. This will leverage the constraint on $\delta m_3^2$ to the level of the current constraints on $\delta m_{1,2}^2$. On the other hand, in the more optimistic scenario where most potential neutrino sources reside farther away, for example at $z \sim 1$, as suggested by the star-formation rate \cite{Hopkins:2006bw, Capanema:2020oet}, the bounds would improve: $\delta m_3^2 \lesssim 10^{-17}$ eV$^2$.

In this work, we will use astrophysical neutrino flavor ratios and the decoherence condition $\Phi_j \gg 1$ to probe pseudo-Dirac neutrinos~\cite{Crocker:1999yw,Crocker:2001zs,Keranen:2003xd,Beacom:2003eu,Esmaili:2009fk,Shoemaker:2015qul,Brdar:2018tce}. In this approach, insensitivity to absolute neutrino flux is a double-edged sword. The advantage is that when there exist hierarchies between $\delta m_i^2$, irrespective of spectral feature, capitalizing on the decoherence effect, stringent upper bounds on larger $\delta m_i^2$ can be derived as we will show in this work. The disadvantage is that if all $\delta m_i^2$ ($i=1,2,3$) are of the same order, we will lose the sensitivity since the transition probabilities for all flavors get approximately the same suppression. This is to be compared with studies which consider specific astrophysical sources where only certain ranges of $\delta m_i^2$ for the first few oscillations can be probed and, away from this region, the analysis is limited by absolute flux uncertainty and knowledge of the spectral feature~\cite{DeGouvea:2020ang,Martinez-Soler:2021unz,Rink:2022nvw,Carloni:2022cqz,Dixit:2024ldv}. If the flux from certain source can be determined quite accurately, and in particular, if all $\delta m_i^2$ are of the same order, this latter approach is superior. In the view that only very few astrophysical neutrino sources are identified, we will focus on flavor ratios approach and see how well this will fare. 

The paper is organized as follows: in Section \ref{sec:model}, we will review the parametrization of pseudo-Dirac model at low energy while in Section \ref{sec:probability}, we will discuss the framework to calculate the oscillation probability in this scenario. 
Then, we will present our main results in Section \ref{sec:results} and conclude in Section \ref{sec:conclusions}. In Appendix~\ref{app:mu-damped}, we investigate the effects of varying astrophysical neutrino sources.

\section{Pseudo-Dirac model}\label{sec:model}
If all three SM neutrinos $\nu_{\alpha}$ ($\alpha=e,\mu,\tau$) are
pseudo-Dirac in nature, we need to introduce the corresponding right-handed
neutrinos $\nu_{\alpha}'$. (While we have used
the same index $\alpha$, the flavors of $\nu_{\alpha}'$ are arbitrary since
they do not feel the SM weak interaction.) After the electroweak symmetry breaking,
we can write down the effective neutrino mass term as $\overline{\Psi^{c}}{\cal M}\Psi$
for $\Psi \equiv \left(\nu_{e},\nu_{\mu},\nu_{\tau},\nu_{e}',\nu_{\mu}',\nu_{\tau}'\right)^{T}$
with
\begin{eqnarray}
{\cal M} & = & \left(\begin{array}{cc}
m_{M} & m_{D}\\
m_{D}^{T} & m_{M}'
\end{array}\right),
\end{eqnarray}
where the total lepton number is conserved by the Dirac mass term $m_{D}$ but broken by the Majorana mass terms $m_{M},m_{M}'$. Without loss of generality, we can work in the basis where
the charged lepton Yukawa is real and diagonal.  We define the pseudo-Dirac
scenario as when the matrix entries satisfying
\begin{eqnarray}
\left|m_{M}\right|,\left|m_{M}'\right| & \ll & \left|m_{D}\right|,\label{eq:pseudo-Dirac}
\end{eqnarray}
where the total lepton number is slightly broken.

Let us first look at the Dirac limit, $m_{M},m_{M}'\to 0$ where the total lepton number
is exactly conserved. In this case, $m_{D}$ can be diagonalized by
two unitary matrices $U_0$ and $V_0$ as follows
%%%
\begin{eqnarray}
\hat{m} & = & U_0^{T}m_{D}V_0
=\textrm{diag}\left(m_{1},m_{2},m_{3}\right).
\end{eqnarray}
%%%
Defining
%%%
\begin{eqnarray}
{\cal U} & = & \frac{1}{\sqrt{2}}\left(\begin{array}{cc}
U_0 & iU_0\\
V_0 & -iV_0
\end{array}\right),\label{eq:Dirac_limit}
\end{eqnarray}
%%%
${\cal M}$ can be diagonalized as ${\cal U}^{T}{\cal M}{\cal U}=\mathrm{diag}\left(\hat{m},\hat{m}\right)$
where the mass eigenstates are given by
\begin{eqnarray}
\hat{\Psi} & = & {\cal U}^{\dagger}\Psi\equiv\left(\nu_{1},\nu_{2},\nu_{3},\nu'_{1},\nu'_{2},\nu'_{3}\right)^{T}.
\end{eqnarray}
We can also express the flavor eigenstates in term of the mass eigenstates as
\begin{eqnarray}
\Psi & = & {\cal U}\hat{\Psi}=\frac{1}{\sqrt{2}}\left(\begin{array}{c}
U_0\left(\nu+i\nu'\right)\\
V_0\left(\nu-i\nu'\right)
\end{array}\right),
\end{eqnarray}
where $\nu \equiv \left(\nu_{1},\nu_{2},\nu_{3}\right)^{T}$ and $\nu' \equiv \left(\nu'_{1},\nu'_{2},\nu'_{3}\right)^{T}$
pair up to form three Dirac states (\emph{maximal} mixing). Experimentally, only $\nu_{\alpha}$ participate in weak interactions, and hence only $U_0$ can be measured. 

When $m_{M},m_{M}'\neq0$ but the pseudo-Dirac condition \eqref{eq:pseudo-Dirac} is still
satisfied, there will be deviation from eq.~\eqref{eq:Dirac_limit}
and we will have three pairs of pseudo-Dirac mass eigenstates with masses ($j=1,2,3$)
\begin{eqnarray}
\hat{m}_{j} & = & m_{j}-\delta m_{j},\qquad\hat{m}_{j+3}=m_{j}+\delta m_{j}.\label{eq:mass_def}
\end{eqnarray}
A useful Euler parametrization for unitary matrix $\mathcal{U}$ is
given by
\begin{eqnarray}
\mathcal{U} & = & \frac{1}{\sqrt{2}}\left(\begin{array}{cc}
AU_0+B & i\left(AU_0-B\right)\\
CU_0+D & i\left(CU_0-D\right)
\end{array}\right),
\end{eqnarray}
where $U_0$ is a $3\times3$ unitary matrix while the rest of the
$3\times3$ matrices $A$, $B$, $C$ and $D$ are constrained by
$\mathcal{U} \mathcal{U}^{\dagger}= \mathcal{U}^{\dagger}\mathcal{U}=I_{6\times6}$. For example, from $\mathcal{U} \mathcal{U}^{\dagger}=I_{6\times6}$,
we have
\begin{eqnarray}
AA^{\dagger}+BB^{\dagger}=CC^{\dagger}+DD^{\dagger} & = & I_{3\times3},\\
AC^{\dagger}+BD^{\dagger}=CA^{\dagger}+DB^{\dagger} & = & 0.
\end{eqnarray}
In the Dirac limit, $A=I_{3\times3}$, $B=C=0$ and $D=V_0$
and eq. (\ref{eq:Dirac_limit}) is recovered.
Explicitly, we can construct the mixing matrix as
\[
\mathcal{U}=U_{\mathrm{new}}UY,
\]
where
\begin{eqnarray*}
U_{\mathrm{new}} & = & R_{56}R_{46}R_{36}R_{26}R_{16}R_{45}R_{35}R_{25}R_{15}R_{34}R_{24}R_{14},\\
U & = & R_{23}R_{13}R_{12},
\end{eqnarray*}
and

\begin{eqnarray*}
Y & \equiv & \frac{1}{\sqrt{2}}\left(\begin{array}{cc}
I_{3\times3} & iI_{3\times3}\\
I_{3\times3} & -iI_{3\times3}
\end{array}\right),
\end{eqnarray*}
with $R_{ij}$ the complex rotation matrix in the $ij$-plane which
can be obtained from a $6\times6$ identity matrix $I$ by replacing
the $I_{ii}$ and $I_{jj}$ by $\cos\theta_{ij}$, $I_{ij}$ by $e^{-i\phi_{ij}}\sin\theta_{ij}$
and $I_{ji}$ by $e^{i\phi_{ij}}\sin\theta_{ij}$. To recover the
Dirac limit, we set all the angles in $U_{\mathrm{new}}$ (besides
$\theta_{45}$, $\theta_{46}$, $\theta_{56}$) to zero. The new angles
$\theta_{45}$, $\theta_{46}$, $\theta_{56}$ which describe the
mixing among the sterile neutrinos (through $V_0$ in the Dirac limit) are in principle not measurable through the SM interactions.

%\footnote{An explicit parametrization is given in Appendix \ref{app:U_parametrization}.} 
While the deviation from maximal mixing $U_0$ is of the order of the following ratios of matrix entries
%%%
\begin{equation}
    \delta U_0 \sim \frac{\left|m_{M}\right|}{\left|m_{D}\right|},\frac{\left|m'_{M}\right|}{\left|m_{D}\right|},
\end{equation}
%%%
%$\delta U_0\sim\left|m_{M}\right|/\left|m_{D}\right|,\left|m'_{M}\right|/\left|m_{D}\right|$,
the pseudo-Dirac mass squared splittings are of the order of $\delta m_{j}^{2}\sim m_{j}^{2}\delta U_0$.
For example, taking $m_{j}\sim0.1$ eV and $\delta U_0\sim10^{-4}$,
the new mass splitting is $\delta m_{j}^{2} \sim10^{-6}\,\textrm{eV}^{2}$. So, it
is expected that the first discovery of pseudo-Diracness should be
through the new mass splittings. In this study, we will assume maximal
mixing and focus only on the effects of new mass splittings. We will leave
the consideration of deviation from maximal mixing for future study.

\section{Oscillation probability}\label{sec:probability}

The amplitude of a neutrino of flavor state $\nu_{\alpha}$ created
at $t=0$ being detected as $\nu_{\beta}$ at time $t > 0$, $S_{\beta\alpha}\left(t\right)\equiv\left\langle \nu_{\beta}|\nu_{\alpha}\left(t\right)\right\rangle $
satisfies the Schr\"{o}dinger equation
\begin{eqnarray}
i\frac{d}{dt}S\left(t\right) & = & HS\left(t\right).
\label{eq:sch_eq}
\end{eqnarray}
The Hamiltonian in the flavor basis is given by
\begin{eqnarray}
H & = & \mathcal{U}\Delta \mathcal{U}^{\dagger}+V,
\end{eqnarray}
where $\mathcal{U}$ is the leptonic mixing matrix, $\Delta\equiv \textrm{diag}\left(\hat{m}_{1}^{2},\hat{m}_{2}^{2},...\right)/(2E)$ with $E$ the neutrino energy,
and $V$ is the matter potential. 
For the Pseudo-Dirac model discussed in the previous section, assuming that the $\nu'_\alpha$ has no new interactions with the SM, the matter potential for ordinary matter is $V = \mathrm{diag}(V_e - V_n,-V_n,-V_n,0,0,0)$ with 
$V_e = \sqrt{2} G_F n_e$ and $V_n = G_F n_n/\sqrt{2}$ where $G_F$ is the Fermi constant, and, $n_e$ and $n_n$ are the electron and neutron number density, respectively.
The solution to eq.~\eqref{eq:sch_eq} is formally given by
\begin{eqnarray}
S & = & T\exp\left[-i\int_{0}^{t}dt'H\left(t'\right)\right],
\end{eqnarray}
where $T$ denotes time ordering.

For astrophysical neutrinos, even if the matter effect is negligible,
there is still nontrivial time-dependence due to cosmic expansion.
Defining the redshift $z$ as
\begin{eqnarray}
1+z & \equiv & \frac{a_{0}}{a},
\end{eqnarray}
where $a$ is the cosmic scale factor with $a_{0}$ the value today,
we have
\begin{eqnarray}
dt & = & \frac{dt}{da}da=-\frac{dz}{\left(1+z\right){\cal H}},
\end{eqnarray}
where ${\cal H}\equiv\frac{1}{a}\frac{da}{dt}$ is the Hubble expansion
rate which can be expressed in term of current expansion rate ${\cal H}_{0}$,
total matter fraction $\Omega_{m}$ and dark energy fraction $\Omega_{\Lambda}$
as follows 
\begin{eqnarray}
{\cal H}(z) & = & {\cal H}_{0}\sqrt{\Omega_{m}\left(1+z\right)^{3}+\Omega_{\Lambda}}.
\end{eqnarray}
In the above, we have ignored the small radiation density and from the measurements of Planck satellite, we have $\Omega_{m}=0.315$, $\Omega_{\Lambda}=0.685$ and ${\cal H}_{0}=67.4\,\textrm{km\,s}^{-1}\textrm{Mpc}^{-1}$~\cite{Planck:2018vyg}.
Furthermore, the energy of the neutrino $E$
detected at $z=0$ will be equal to $E\left(1+z\right)$ at redshift
$z>0$. With negligible matter effect for astrophysical neutrinos $V=0$, we can write the solution as
\begin{eqnarray}
S_{\beta\alpha} & = & \sum_{j}{\cal U}_{\beta j}{\cal U}_{\alpha j}^{*}
e^{-i\frac{m_{j}^{2}}{2E} L_{\textrm{eff}}},
\end{eqnarray}
where we have defined the effective distance as
\begin{eqnarray}
L_{\textrm{eff}} & \equiv & c\int_{0}^{z}\frac{dz'}{\left(1+z'\right)^{2}{\cal H}(z')},
\end{eqnarray}
with the speed of light $c$ shown explicitly due to the unit of ${\cal H}_{0}$ that we have chosen. 
For $z < 1$, we can Taylor expand in $z$ and obtain at the leading order 
\begin{equation}
L_{\textrm{eff}} \simeq \frac{cz}{{\cal H}_{0}}
= 4.4 \,\textrm{Mpc}
\left(\frac{z}{0.001}\right)
\left(\frac{67.4\,\textrm{km\,s}^{-1}\textrm{Mpc}^{-1}}{{\cal H}_{0}}\right),
\end{equation}
where the corrections are ${\cal O}(z^2)$.
Finally, the transition probability is given by the Born rule $P\left(\nu_{\alpha}\to\nu_{\beta}\right) \equiv \left|S_{\beta\alpha}\right|^{2}$.

Assuming the new angles are negligible but new mass splittings are
relevant, we have the oscillation probability
\begin{eqnarray}
P\left(\nu_{\alpha}\to\nu_{\beta}\right) 
 & = & \left|\sum_{j=1}^{3}U_{0,\beta j}U_{0,\alpha j}^{*}e^{-i\frac{m_{j}^{2}}{2E} L_{\textrm{eff}}}\cos\left(\Phi_j\right)\right|^{2},
\end{eqnarray}
where we have defined
\begin{equation}
    \Phi_j \equiv \frac{\delta m_{j}^{2}}{4E}L_{\textrm{eff}},
\end{equation}
and used $\delta m_{j}^{2}\simeq4m_{j}\delta m_{j}$ from the definitions in eqs.~\eqref{eq:new_squared_mass_splitting} and \eqref{eq:mass_def}.
Astrophysical neutrino sources are far enough such that we can average out all the standard oscillations involving $m_{j}^{2}-m_{k}^{2}\neq0$
and obtain
\begin{eqnarray}
P\left(\nu_{\alpha}\to\nu_{\beta}\right) & \simeq & \sum_{j=1}^{3}\left|U_{0,\beta j}U_{0,\alpha j}^{*}\right|^{2}\cos^{2}\left(\Phi_j\right).
\end{eqnarray}
If all $\delta m_{j}^{2}$ are also large enough 
%%%
\begin{equation}
    \Phi_j
    = 10 \left(\frac{\delta m_j^2 }{1.7\times 10^{-12}\,\textrm{eV}^2}\right)
    \left(\frac{L_\textrm{eff}}{15\,\textrm{kpc}}\right)
    \left(\frac{100\,\textrm{TeV}}{E}\right) \gg 1,\label{eq:condition}
\end{equation}
%%% 
such that we can average out these new oscillations, we obtain
\begin{eqnarray}
P\left(\nu_{\alpha}\to\nu_{\beta}\right) & \simeq & \frac{1}{2}\sum_{j=1}^{3}\left|U_{0,\beta j}U_{0,\alpha j}^{*}\right|^{2}.\label{eq:democratic}
\end{eqnarray}
In general, eq.~\eqref{eq:democratic} changes the flux normalization for all flavors.
However, if the absolute flux is unknown, it will be challenging to distinguish it from the standard scenario.

On the other hand, using neutrino flavor ratios which do not depend on absolute flux, one can distinguish between the following six cases: Case 1, Case 2, Case 3, Case 12, Case 23, and Case 13 where Case $j$ is defined as
$\delta m_{k\neq j}^2=0$ and $\Phi_j\gg1$ with oscillation probability
\begin{eqnarray} \label{probj}
%P\left(\nu_{\alpha}\to\nu_{\beta}\right) 
P_{\alpha\beta}^j & \equiv & 
\frac{1}{2}\left|U_{0,\beta j}U_{0,\alpha j}^{*}\right|^{2}
+ \sum_{k\neq j}\left|U_{0,\beta k}U_{0,\alpha k}^{*}\right|^{2},
\end{eqnarray}
and Case $jk$ is defined as
$\delta m_{l\neq \{j,k\}}^2=0$ and $\Phi_{j,k}\gg1$ with oscillation probability
\begin{eqnarray} \label{probjk}
%P\left(\nu_{\alpha}\to\nu_{\beta}\right) 
P_{\alpha\beta}^{jk} & \equiv & 
\frac{1}{2}\left|U_{0,\beta j}U_{0,\alpha j}^{*}\right|^{2}
+ \frac{1}{2}\left|U_{0,\beta k}U_{0,\alpha k}^{*}\right|^{2}
+ \sum_{l\neq \{j,k\}}\left|U_{0,\beta l}U_{0,\alpha l}^{*}\right|^{2}.
\end{eqnarray}
By $\delta m_{k\neq j}^2=0$, we mean $\delta m^2_{k\neq j} \ll \delta m_{j}^2$ such that oscillations involving $\delta m_{k\neq j}^2$ have not developed. 
For example, considering the most distant identified source of astrophysical neutrinos PKS 1424+240 of 2.6 Gpc together with $E = 1\,{\rm TeV}$~\cite{Paiano:2017pol,Carloni:2022cqz}, we need $\delta m^2_{k\neq j} \ll 10^{-20}\,{\rm eV}^2$ such that the oscillation involving $\delta m^2_{k\neq j}$ has not developed. Since very few astrophysical neutrino sources have been identified, the best we can currently do is to make assumptions based on the six possible scenarios mentioned above. 

The relevant dataset of High-Energy Starting Events (HESE)~\cite{IceCube:2020wum}, from which neutrino flavors can be distinguished by IceCube, includes events with energies above 60 TeV and up to approximately 10 PeV. According to Eq.~\eqref{eq:condition}, the corresponding sensitivity range of pseudo-Dirac mass splittings for our analysis is shown for each case in Table~\ref{tab:6_cases}. For example, in Case 3, where the expected regions for the standard and pseudo-Dirac scenarios do not overlap significantly (see next section), if the data remains consistent with the standard scenario at a given confidence level, we can place an upper bound on the mass splitting, $\delta m_3^2 \lesssim 10^{-10}~\text{eV}^2$, assuming $\delta m_{1,2}^2 = 0$. This bound corresponds to the decoherence condition set by the upper end of the energy window, approximately 10 PeV. More stringent upper bounds, however, can be obtained by using lower-energy neutrinos around 100 TeV, for which the decoherence condition in Eq.~\eqref{eq:condition} implies $\delta m_3^2 \lesssim 10^{-12}~\text{eV}^2$. Statistically, this latter bound is expected to dominate (and is the one quoted in Table~\ref{tab:6_cases}), as lower-energy neutrinos yield higher event statistics~\cite{IceCube:2020wum}.

\begin{table} 
\begin{tabular}{|c|c|} 
\hline 
Case & Pseudo-Dirac mass splitting\tabularnewline
\hline 
\hline 
Case 1 & $\delta m_{1}^{2} \gtrsim 10^{-12}$ ${\rm eV}^{2}$, $\delta m_{2,3}^{2}= 0$
 \tabularnewline
\hline 
Case 2 & $\delta m_{2}^{2} \gtrsim 10^{-12}$ ${\rm eV}^{2}$, $\delta m_{1,3}^{2}= 0$
 \tabularnewline
\hline 
Case 3 & $\delta m_{3}^{2} \gtrsim 10^{-12}$ ${\rm eV}^{2}$, $\delta m_{1,2}^{2}= 0$
 \tabularnewline
\hline 
Case 12 & $\delta m_{1,2}^{2} \gtrsim 10^{-12}$ ${\rm eV}^{2}$, $\delta m_{3}^{2} = 0$
 \tabularnewline
\hline 
Case 23 & $\delta m_{2,3}^{2} \gtrsim 10^{-12}$ ${\rm eV}^{2}$, $\delta m_{1}^{2} = 0$
 \tabularnewline
\hline 
Case 13 & $\delta m_{1,3}^{2} \gtrsim 10^{-12}$ ${\rm eV}^{2}$, $\delta m_{2}^{2} = 0$ \tabularnewline
\hline 
\end{tabular}

\caption{The cases with their corresponding pseudo-Dirac mass splitting ranges are valid for any astrophysical neutrinos with
$E \sim 100\ {\rm TeV}$ from extragalactic sources i.e $L>15\,{\rm kpc}$. The meaning of $\delta m^2_j = 0$ can be taken to be $\delta m^2_j \ll 10^{-20}\,{\rm eV}^2$ considering the most distant identified astrophysical neutrino source (see text for discussion).}\label{tab:6_cases}
\end{table}

\subsection{Neutrino flavor ratios}

In the standard scenario, the oscillation probability between an initial flavor $\nu_\alpha$ to a final flavor $\nu_\beta$ is given by
\begin{equation}
    P_{\alpha \beta}^{std}= \sum_{i=1}^3 |U_{\alpha i}|^2 |U_{\beta i}|^2.
\end{equation}
Therefore, for a flavor composition $(f_{e,S},f_{\mu,S},f_{\tau,S})$ at the source $S$, we can compute the final flavor at the Earth as 
\begin{equation}
    f_{\beta, \oplus}=\sum_{\alpha=e,\mu,\tau} P_{\alpha \beta}^{std} f_{\alpha,S}.
\end{equation}

In the presence of averaged-out pseudo-Dirac oscillations, we must change $P_{\alpha \beta}^{std}$ to $P_{\alpha \beta}^{j}$ or $P_{\alpha \beta}^{jk}$, where the indeces $j$ and $jk$ correspond to the pseudo-Dirac scenarios described by eq.~\eqref{probj} and eq.~\eqref{probjk}. Therefore, for such cases, the final fractions of active flavors $\nu_\beta$ are given by $f_{\beta, \oplus}^j=\sum_{\alpha} P_{\alpha \beta}^{j} f_{\alpha,S}$ and $f_{\beta, \oplus}^{jk}=\sum_{\alpha} P_{\alpha \beta}^{jk} f_{\alpha,S}$, respectively. However, due to the finite conversion probability between active and sterile states, the total flux of active neutrinos will not be conserved from production to detection. Therefore, to determine the detected fractions, we need to normalize $f_{\beta, \oplus}^{j}$ and $f_{\beta, \oplus}^{jk}$:
\begin{equation} \label{fdet}
    f_{\beta, \oplus}^{j,\text{det}}=\frac{f_{\beta, \oplus}^{j}}{f_{e, \oplus}^{j}+f_{\mu, \oplus}^{j}+f_{\tau, \oplus}^{j}} \hspace{0.5cm} \text{and} \hspace{0.5cm} f_{\beta, \oplus}^{jk,\text{det}}=\frac{f_{\beta, \oplus}^{jk}}{f_{e, \oplus}^{jk}+f_{\mu, \oplus}^{jk}+f_{\tau, \oplus}^{jk}}.
\end{equation}

Next, we will focus on the aforementioned six cases and evaluate the sensitivity of IceCube and IceCube-Gen2 in constraining/distinguish them.

\section{Results}\label{sec:results}

HE astrophysical neutrinos are generated in cosmic particle accelerators when accelerated protons collide with matter ($pp$ interactions) and radiation ($p\gamma$ interactions) \cite{Margolis:1977wt, Stecker:1978ah, Mucke:1999yb, Kelner:2006tc, Hummer:2010vx}. These collisions produce mesons, particularly charged pions ($\pi^\pm$), which then decay into neutrinos. The decay chain $\pi^- \rightarrow \mu^- + \Bar{\nu}_\mu \rightarrow e^- + \Bar{\nu}_e + \nu_\mu + \Bar{\nu}_\mu$, along with its charge conjugate reaction, produces two muon neutrinos for every electron neutrino: $(1,2,0)_S$, where the subscript $S$ denotes the initial proportions at the source\footnote{Neutrinos and antineutrinos cannot be distinguished at the IceCube detector. The only exception is the Glashow Resonance \cite{Glashow:1960zz, Bhattacharya:2011qu, Huang:2023yqz, Liu:2023lxz}, but it suffers from a lack of statistics, with only one candidate event detected so far~\cite{IceCube:2021rpz}.}. It is also possible that the decay chain described above is not complete, as the muon may lose energy due to synchrotron emission before decaying. As a result, this will generate the $\mu$-damped scenario, with the proportions $(0,1,0)_S$ \cite{Rachen:1998fd, Kashti:2005qa, Kachelriess:2007tr, Hummer:2010ai, Winter:2014pya}. In this section, however, we assume neutrinos are produced only by the complete chain of $\pi^\pm$ decay and reach the Earth with averaged-out pseudo-Dirac oscillations, as described by eq.~\eqref{probj} and eq.~\eqref{probjk}, and compare these scenarios with the standard expectations for both $\pi$-decay and $\mu$-damped. More information regarding the $\mu$-damped production combined with the pseudo-Dirac scenario is reserved for the Supplemental Material.

\begin{figure*}[htb!]
\includegraphics[width=0.99\textwidth]{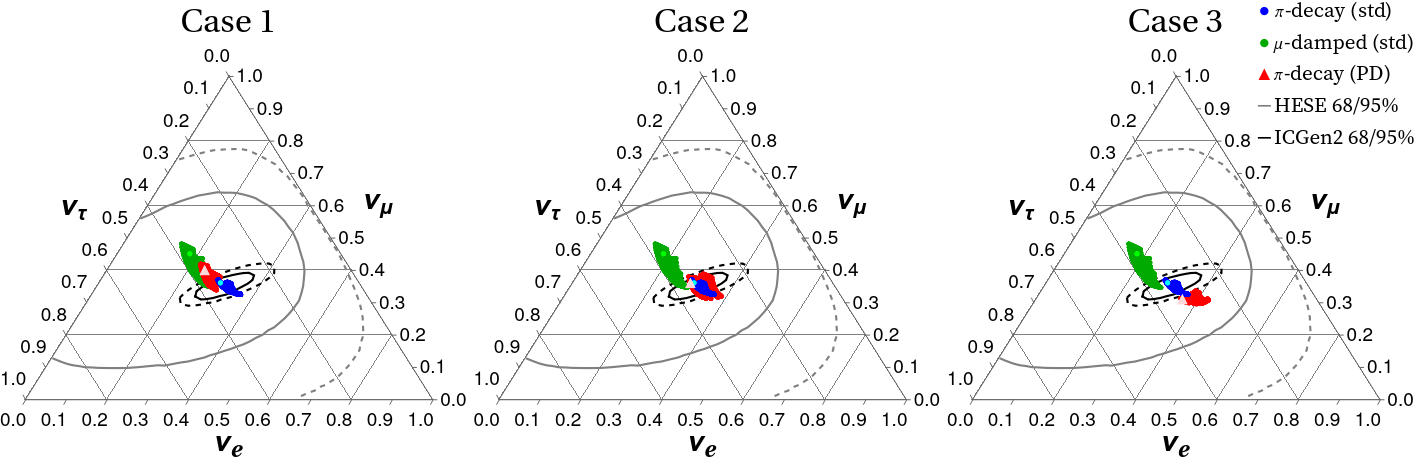}
  \caption{Expected regions for the flavor composition of HE astrophysical neutrinos, considering the $3 \sigma$ interval for the oscillation parameters in \cite{deSalas:2020pgw}, valid for both normal and inverted mass orderings. The standard (std) expectation for $\pi$-decay is shown in blue, the standard $\mu$-damped is shown in green, and the various scenarios for pseudo-Dirac neutrinos are shown in red. 
  In Case $j$, oscillations corresponding to pseudo-Dirac squared mass splitting $\delta m^2_j$ have been averaged out.
  See the text for more details about the different pseudo-Dirac scenarios.}
\label{money_Fig_j}
\end{figure*}

\begin{figure*}[htb!]
\includegraphics[width=0.99\textwidth]{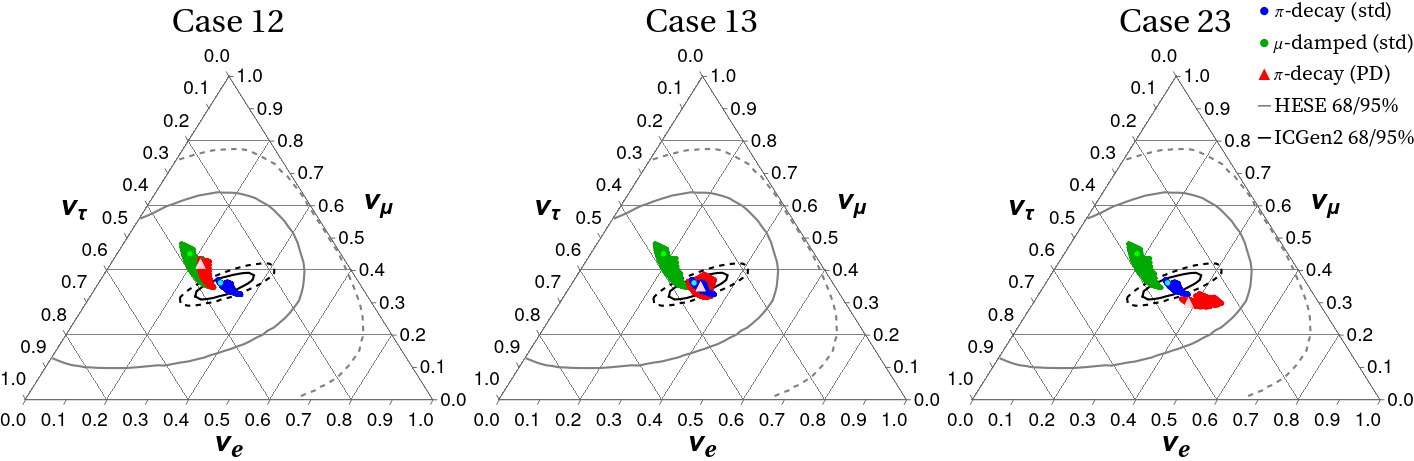}
  \caption{Same as fig.~\ref{money_Fig_j}, but for pseudo-Dirac scenarios named Case $jk$. 
  }
\label{money_Fig_jk}
\end{figure*}

During propagation from the sources to the Earth, neutrino oscillations modify the initial flavor proportions. In the standard vacuum oscillation scenario, assuming the absence of matter effects \cite{Dev:2023znd} and pseudo-Dirac mass splittings, the $\pi^\pm$ production channel results in flavor equipartition at Earth: $(1:2:0)_S \rightarrow (1:1:1)_\oplus$. To be more precise, in fig.~\ref{money_Fig_j} and fig.~\ref{money_Fig_jk}, we show the allowed region for the flavor composition at the Earth after considering the $3 \sigma$ uncertainties of the oscillation parameters~\cite{deSalas:2020pgw} where there is no perceptible distinction between normal and inverted mass orderings.
Due to the smallness of $\sin\theta_{13}$, the effect of leptonic CP phase $\delta_{\rm CP}$ on allowed region is small and for this reason and to be as conservative as possible, we do not take into account the correlation between $\theta_{23}$ and $\delta_{\rm CP}$.
Observe that the $\pi$-decay (blue region) using the best fit oscillation parameters of ref.~\cite{deSalas:2020pgw}, in the standard case, is given by $(0.3,0.36,0.34)_\oplus$, which lies very close to the benchmark of total flavor equipartition. On the other hand, for the $\mu$-damped case (green region), the best fit is $(0.18,0.45.0.37)_\oplus$. 
Also represented in fig.~\ref{money_Fig_j} and fig.~\ref{money_Fig_jk} are the regions corresponding neutrino flavor at the Earth in the presence of averaged-out pseudo-Dirac oscillation as described by eq.~\eqref{probj} and eq.~\eqref{probjk} (red regions). We investigate all possible cases listed in the section~\ref{sec:probability}. 

We observe that Case 3 and Case 23 stand out as the most distinct from the standard scenarios, even when considering the current uncertainties in the oscillation parameters and production processes ($\pi$-decay or $\mu$-damped).
What makes the regions corresponding to Cases 3 and 23 so distinct from the standard scenarios is that by decreasing the amount of $\nu_3$ in the flux, we automatically reduce the proportion of $f_{\mu,\oplus}$\footnote{This fraction here should be undertood as the detected fraction $f^{\text{det}}$.} (as well as $f_{\tau,\oplus}$, but in a smaller quantity), due to the larger mixing of this propagation eigenstate with $\nu_\mu$. Applying the same reasoning, if some of the $\nu_1$ eigenstates disappear (Case 1 or Case 12), $f_{e,\oplus}$ decreases, and the expected region shifts closer to the typical $\mu$-damped scenario. For Case 13, however, the decrease $f_{\mu, \oplus}$ due to the disappearance of $\nu_3$ is balanced by the decrease in $f_{e, \oplus}$ resulting from the disappearance of $\nu_1$, making the final region compatible with the standard $\pi$-decay. Moreover, $\nu_2$ has roughly equal mixing with all flavors, so its disappearance (Case 2) has a smaller effect. These results hold true for both normal and inverted mass orderings.

Fig.~\ref{money_Fig_j} and fig.~\ref{money_Fig_jk} also shows the $68\%$ and $95\%$ C.L. contours corresponding to the IceCube analysis \cite{IceCube:2020fpi} of the flavor composition of the HE astrophysical neutrinos using the HESE sample \cite{IceCube:2020wum}. HESE is an all-flavor, all-sky search using 7.5 years of IceCube data for neutrinos with energies above 60 TeV. Observe that we cannot draw any conclusions from the current HESE bounds, which are still too weak to discriminate even between the standard scenarios. Nevertheless, future detectors, such as IceCube-Gen2 \cite{IceCube-Gen2:2020qha}, with more statistics and improved capabilities, might significantly enhance the sensitivity to the flavor composition. In fig.~\ref{money_Fig_j} and fig.~\ref{money_Fig_jk} we also show the $68\%$ and $99\%$ projected sensitivities of IceCube-Gen2 after 10 years of observations. The projection of IceCube-Gen2 is taken from \cite{IceCube-Gen2:2023rds} and assumes the best fit value to be  $(0.3,0.36,0.34)_\oplus$, compatible with standard $\pi$-decay. Observe that with IceCube-Gen2, we can start to distinguish between the standard scenario and some pseudo-Dirac scenarios (Case 3 and Case 23), even if the production process ($\pi$-decay and/or $\mu$-damped) remains uncertain at that time. In Table~\ref{tab:pion_source}, we summarize the best-fit flavor fractions for each of the cases illustrated in fig.~\ref{money_Fig_j} and fig.~\ref{money_Fig_jk}, normalized as in Eq.~\eqref{fdet}.
In Appendix~\ref{app:mu-damped}, we show the results assuming $\mu$-damped channel and combination with $\pi$-decay and neutron decay channels of varying proportions for pseudo-Dirac scenario and in those cases, the differences between the standard and pseudo-Dirac scenarios are less pronounced.

\renewcommand{\arraystretch}{0.9} % vertical padding

\begin{table}[h!]
\centering

\begin{tabular}
{|>{\centering\arraybackslash}p{4cm}|>{\centering\arraybackslash}p{6cm}|}
\hline
\multicolumn{2}{|c|}{\textbf{At the Source}} \\
\hline
$\pi$-decay & \textbf{($f_{e,S}$, $f_{\mu,S}$, $f_{\tau,S}$)}$ \ =\left( \dfrac{1}{3}, \dfrac{2}{3}, 0 \right)$ \\
\hline
\multicolumn{2}{|c|}{\textbf{At the Earth}} \\
\hline
\textbf{Scenario} & ($f_{e,\oplus}^{\text{det}}$, $f_{\mu,\oplus}^{\text{det}}$, $f_{\tau,\oplus}^{\text{det}}$) \\
\hline
Std. oscillations & (0.30, 0.36, 0.34) \\
\hline
Case 1 & (0.24, 0.40, 0.36) \\
\hline
Case 2 & (0.29, 0.36, 0.35) \\
\hline
Case 3 & (0.36, 0.31, 0.33) \\
\hline
Case 12 & (0.22, 0.42, 0.36) \\
\hline
Case 13 & (0.32, 0.35, 0.33) \\
\hline
Case 23 & (0.38, 0.30, 0.32) \\
\hline
\end{tabular}

\caption{Flavor composition at Earth for different scenarios using the best fit oscillation parameters of ref.~\cite{deSalas:2020pgw}, assuming pion decay at the source.} \label{tab:pion_source}
\end{table}

Finally, the more precise determination of several oscillation parameters in the next decade will contribute to astrophysical flavor measurements and enhance the discriminatory power of future detectors 
and taking into account the projected increase in precision in measurements of mixing parameters $\theta_{12}$ and $\theta_{23}$ in year 2040, the allowed regions practically shrink to best-fit points indicated by colored circles and triangle in fig.~\ref{money_Fig_j} and fig.~\ref{money_Fig_jk}~\cite{Song:2020nfh}.

\section{Conclusions}\label{sec:conclusions}

In this work, we have shown that using astrophysical neutrino flavor ratios, it is possible to distinguish between six possible cases of pseudo-Dirac neutrino spectra where there exists hierarchies among the new squared mass splitting. 
Our main finding is that the most sensitive cases turn out to involve the pseudo-Dirac mass splitting of the third mass eigenstates $\delta m_3^2$ which currently has the weakest constraint of the order $10^{-5}\,{\rm eV}^2$.
For the cases $\delta m_{1,2}^2 =0$ or $\delta m_1^2 = 0$, a conservative upper bound of order $\delta m_3^2 \lesssim 10^{-12}\,\textrm{eV}^2$ can be obtained considering astrophysical neutrinos of energies $E \sim 100$ TeV coming from a distance greater than the size of our galaxy and assuming $\pi$-decay at neutrino sources with IceCube-Gen2 experiment. If nature is kind to us, we might detect the first positive evidence of pseudo-Dirac neutrino with $\delta m_3^2 \sim 10^{-12}\,\textrm{eV}^2$ with astrophysical neutrino flavor ratio.
From eq.~\eqref{eq:condition}, if extragalatic neutrino sources are determined, the bound will greatly improve, up to five orders of magnitude if the sources are of the distance of Gpc. The benefit of using flavor ratios is its insensitivity to absolute neutrino flux uncertainty while the main caveat is that if all pseudo-Dirac mass splittings are of the same order, one will obtain an overall reduction of flux, making this method futile. 
As more astrophysical neutrino sources are identified, one can combine the flavor ratio method together with direct measurements to derive the best constraints on pseudo-Dirac neutrinos, exploring completely the window of pseudo-Dirac leptogenesis proposed in ref.~\cite{Fong:2020smz}.

\section*{Acknowledgments}
 Y.P. acknowledges support by Fundação de Amparo à Pesquisa do Estado de São Paulo (FAPESP) Contracts No.  2023/10734-3 and No. 2023/01467-1, and by the
National Council for Scientific and Technological
Development (CNPq) Grant No. 151168/2023-7. C.S.F. acknowledges support by FAPESP Contracts No. 2019/11197-6 and No. 2022/00404-3 and Conselho Nacional de Desenvolvimento Científico e Tecnológico (CNPq) under Contracts No. 407149/2021-0 and No. 304917/2023-0.
\ \

\textbf{Note Added:} A recent work \cite{Dev:2024yrg} has also considered effects of pseudo-Dirac neutrinos on astrophysical neutrino flavor ratios and showed that if there is an overdensity of cosmic neutrino background by a factor of $10^4$ around the Earth, matter effect can be relevant. Since gravitational effect can only give rise to order of one enhancement, further new physics is required to achieve this large enhancement. To be conservative, we will assume no such large overdensity and ignore the neutrino matter effect.

\appendix
\section{Appendix}

\subsection*{$\mu$-damped production and the combination with $\pi$-decay and neutron decay production}
\label{app:mu-damped}

\begin{figure*}[htb!]
\includegraphics[width=0.99\textwidth]{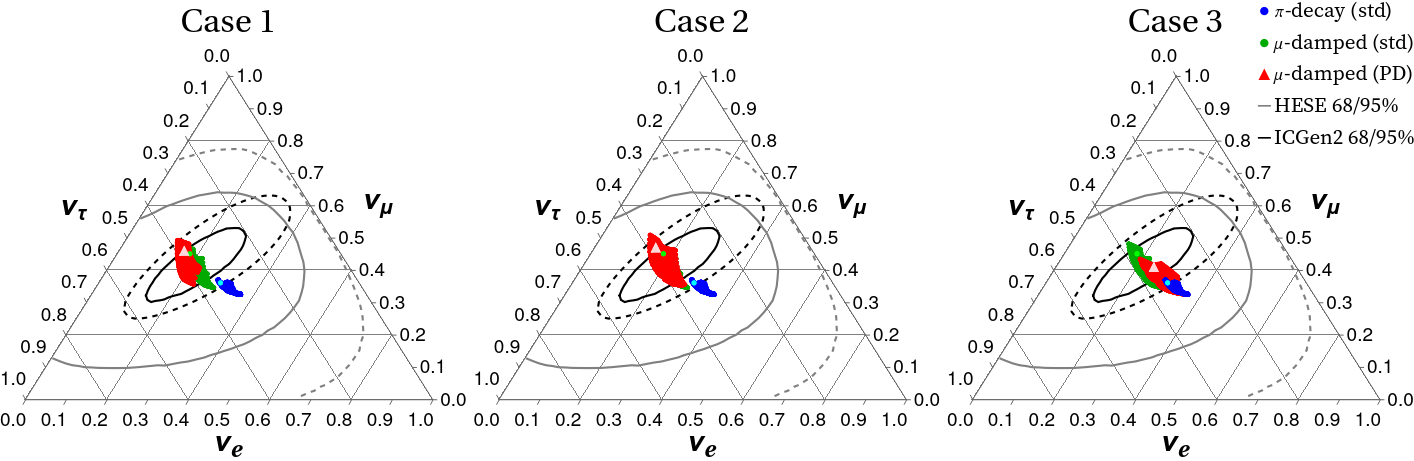}
  \caption{Same as fig.~\ref{money_Fig_j} in the main text, but with the assumption that pseudo-Dirac neutrinos are produced via the $\mu$-damped channel and corresponding projection of IceCube-Gen2 is taken from \cite{IceCube-Gen2:2023rds}.
  }
\label{money_Fig_j_app}
\end{figure*}

\begin{figure*}[htb!]
\includegraphics[width=0.99\textwidth]{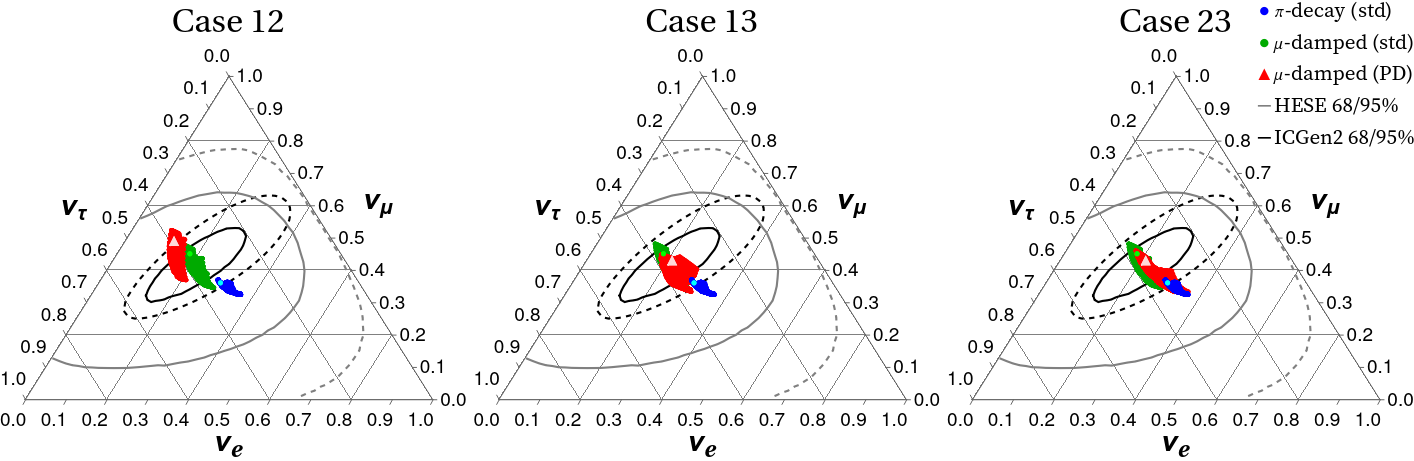}
  \caption{Same as fig.~\ref{money_Fig_jk}  in the main text, but with the assumption that pseudo-Dirac neutrinos are produced via the $\mu$-damped channel and corresponding projection of IceCube-Gen2 is taken from \cite{IceCube-Gen2:2023rds}. 
  }
\label{money_Fig_jk_app}
\end{figure*}

We first show the results for the combination of $\mu$-damped production with the pseudo-Dirac scenario. We highlight that only in Case 12 the pseudo-Dirac scenario appears as clearly distinct from standard scenarios, although it will still be hard to discriminate even after 10 years of IceCube-Gen2 operation.
As for the pseudo-Dirac scenarios involving $\nu_3$ mass eigenstates (Case 3, Case 13 and Case 23) assuming $\mu$-damped sources, it can result in confusion with the standard scenario assuming $\pi$-decay sources, see figs.~\ref{money_Fig_j_app} and \ref{money_Fig_jk_app}. 
In Table~\ref{tab:muon_damped}, we list the best-fit flavor fractions for each of the cases illustrated in fig.~\ref{money_Fig_j_app} and fig.~\ref{money_Fig_jk_app}.

\begin{table}[h!]
\centering

\begin{tabular}
{|>{\centering\arraybackslash}p{4cm}|>{\centering\arraybackslash}p{6cm}|}
\hline
\multicolumn{2}{|c|}{\textbf{At the Source}} \\
\hline
$\mu$-damped & \textbf{($f_{e,S}$, $f_{\mu,S}$, $f_{\tau,S}$)}$ \ =\left( 0, 1, 0 \right)$ \\
\hline
\multicolumn{2}{|c|}{\textbf{At the Earth}} \\
\hline
\textbf{Scenario} & ($f_{e,\oplus}^{\text{det}}$, $f_{\mu,\oplus}^{\text{det}}$, $f_{\tau,\oplus}^{\text{det}}$) \\
\hline
Std. oscillations & (0.18, 0.45, 0.37) \\
\hline
Case 1 & (0.16, 0.46, 0.38) \\
\hline
Case 2 & (0.15, 0.47, 0.38) \\
\hline
Case 3 & (0.24, 0.41, 0.35) \\
\hline
Case 12 & (0.12, 0.49, 0.39) \\
\hline
Case 13 & (0.24, 0.42, 0.34) \\
\hline
Case 23 & (0.23, 0.41, 0.36) \\
\hline
\end{tabular}

\caption{Flavor composition at Earth for different scenarios using the best fit oscillation parameters of ref.~\cite{deSalas:2020pgw}, assuming muon damped at the source.} \label{tab:muon_damped}
\end{table}

\begin{figure*}[htb!]
\includegraphics[width=0.99\textwidth]{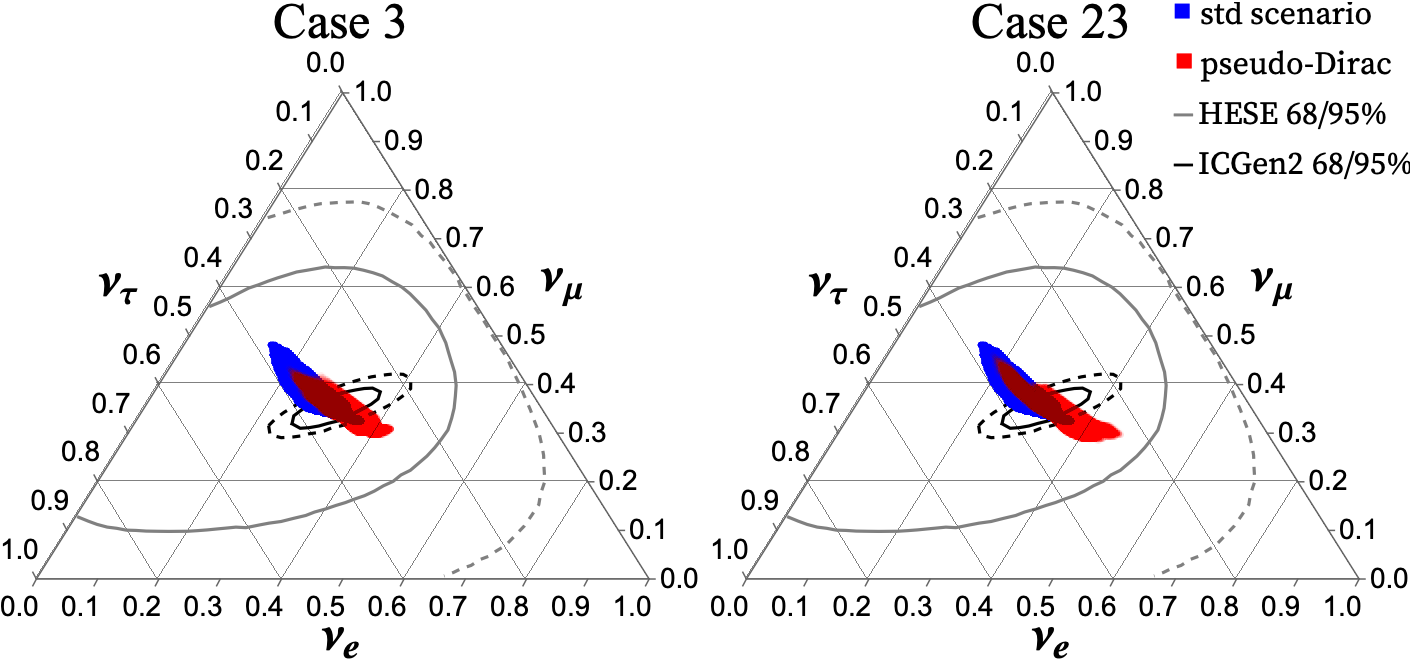}
  \caption{Expected regions for the flavor composition of high-energy astrophysical neutrinos in the standard scenario (blue) and the pseudo-Dirac scenario (red). In both cases, we vary the oscillation parameters and consider two production mechanisms (pion decay and muon-damped) of varying proportions to obtain the allowed regions. Note that the standard and pseudo-Dirac scenarios largely overlap, except for the lower portion of the red region (originating from pion-decay sources emitting pseudo-Dirac neutrinos; see figs.~\ref{money_Fig_j} and \ref{money_Fig_jk}), which could, in principle, be distinguished by future experiments such as IceCube-Gen2.}
\label{money_Fig_full}
\end{figure*}

In fig.~\ref{money_Fig_full}, we show ternary plots for the allowed regions in both the standard and pseudo-Dirac scenarios for the case in which $\pi$-decay and $\mu$-damped contributions, in varying proportions, are simultaneously present. To show how this can affect the two most interesting cases (Cases 3 and 23), we present the analysis for them. The result is that corresponding regions shown in figs.~\ref{money_Fig_j}-\ref{money_Fig_jk_app} merge significantly. The blue region corresponds to a mixture of $\mu$-damped and $\pi$-decay contributions in varying proportions, assuming the standard scenario. It largely overlaps with the red region, which represents the same mixture but in the presence of pseudo-Dirac neutrinos (the purple or dark red region indicates the overlap between the blue and light red areas).

\begin{figure*}[htb!]
\includegraphics[width=0.99\textwidth]{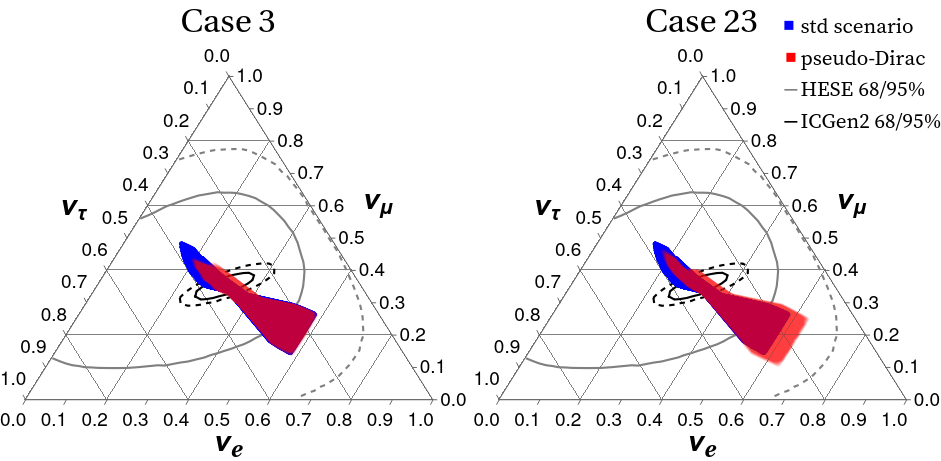}
  \caption{Same as Fig.~\ref{money_Fig_full}, but in addition to pion decay and muon-damped production, we include neutron decay as a possible production mechanism. Neutron decay leads to source flavor ratios of the form $(1, 0, 0)_S$ and corresponds to flavor ratios at Earth approximately $(0.55, 0.17, 0.28)_\oplus$, which pushes the allowed region toward the lower-right vertex of the flavor triangle~\cite{Song:2020nfh}.}
\label{money_Fig_full_a}
\end{figure*}

In fig.~\ref{money_Fig_full_a}, we add to the set of possible production processes the neutron decay, which is also a benchmark scenario often quoted in the literature. Neutron decay produces neutrinos with source flavor ratios $(1, 0, 0)_S$, which evolve into an approximate composition of $(0.55, 0.17, 0.28)_\oplus$ at Earth, thereby shifting the allowed region in the flavor triangle toward its lower-right corner in both standard and pseudo-Dirac scenarios~\cite{Song:2020nfh}.

\bibliography{reference}

\end{document}